\documentclass[pra,aps,twocolumn,10pt]{revtex4-1}
\usepackage{amssymb,amsmath,graphicx}

\begin{document}

\title{Nonlinear lossy light bullets in self-focusing media with nonlinear absorption}

\author{Miguel A. Porras}

\affiliation{Grupo de Sistemas Complejos, ETSIME, Universidad Polit\'ecnica de Madrid, Rios Rosas 21, 28003 Madrid, Spain}

\begin{abstract}
We review the properties of nonlinear, multidimensional localized waves whose stationary propagation is sustained by a dynamic
equilibrium between self-focusing and nonlinear losses. Their finite-energy versions preserve light bullet behavior well-beyond the characteristic diffraction or dispersion distances, and rebuild after obstacles. There exists a preferential
lossy light bullet with maximum intensity and losses, defined solely by the optical properties of the medium, which is the most 
stable, non-conical localized wave supported by a medium with self-focusing nonlinearity and nonlinear losses. This preferential lossy light bullet acts an as attractor during self-focusing of Gaussian-like wave packets when collapse is halted by nonlinear absorption, which can explain relevant properties of the observed light filament dynamics in media with anomalous dispersion.
\end{abstract}

\maketitle

\section{Introduction}

\noindent In this article we review the properties of the so-called nonlinear lossy light bullets (LLBs). They constitute a family of localized and non-diffractive light wave packets in homogeneous, isotropic, nonlinear media 
that results from a dynamic balance between energy dissipation and self-focusing \cite{PORRAS3,PORRAS5}. Their properties differ substantially
from those of other, well-known families of non-diffractive, nonlinear light bullets, as solitary, conical and dissipative bullets. LLBs are substantially nonlinear and multidimensional waves (2D and 3D), and feature a rather complex spatiotemporal structure. They survive to nonlinear absorption and rebuild after obstacles, but these properties do not result from a peculiar geometry, as in conical light bullets \cite{DURNIN,BOUCHAL,DUBIETIS1,DUBIETIS2,PORRAS1,PORRAS2}, or on a balancing gain, as in dissipative light bullets \cite{AKHMEDIEV,GRELU,ANKIEWICZ}. The stationary propagation of LLBs against nonlinear losses is a result of the refilling effect of self-focusing. Self-focusing establishes an energy flux from an energy reservoir in the light bullet periphery towards the nonlinearly absorbed central core of the bullet.
Further, nonlinear losses make the propagation of these LLBs more robust against perturbations.
In each nonlinear medium, there exists a LLB with maximum intensity and maximum losses, defined by the optical properties of the medium solely, that
exhibits maximum stability properties, and that act as an attractor of the self-focusing dynamics with nonlinear losses. The existence of this
attractor can explain many facts of the self-focusing, collapse and filamentation dynamics when collapse is arrested by nonlinear losses
\cite{HENZ,MOLL,SKUPIN}.

Along this paper, these properties and their relevance in experiments are reviewed. In Sections \ref{LLBS} and \ref{STRUCTURE} we introduce LLBs as localized and stationary solutions of the two or three dimensional nonlinear Schr\"odinger equation with self-focusing nonlinearity and nonlinear losses, and describe their rather complex structure, formed by a narrow peak, surrounded by a dissipative shell and an energy reservoir. In the LLB with maximum intensity and maximum losses supported by the medium, the dissipative shell extends infinitely far from the bullet center. The propagation properties of the physically realizable, finite-energy versions of LLBs are studied in Section \ref{FINITE}, where it is seen that truncated LLBs can propagate as light bullets for hundreds of diffraction lengths. Section \ref{SELF} describes the self-reconstruction property of LLBs after obstacles.
Special attention is paid to the difficult question of the stability of LLBs. In Section \ref{ATTRACTIVE} we show that the LLB of maximum intensity
and losses tends to be spontaneously formed in the collapse of standard wave packets with finite energy arrested by nonlinear losses
\cite{PORRAS3,PORRAS4}. The dynamics towards the formation, and relaxation from this LLB attractor is seen to reproduce many features of the
filamentation dynamics of monochromatic light (in the two-dimensional case), and of the filamentation dynamics of ultrashort pulses in media with
anomalous dispersion (in the three-dimensional case), as the filament intensity, the particularly long segments and revivals in the form of short
bursts. The stability properties of LLBs are studied in Section \ref{ANALYSIS} by means of a linearized instability analysis. As expected from its
attractive property, the LLB of maximum intensity and losses turns out to be the most stable among all LLBs.

\section{Lossy light bullets in self-focusing media with nonlinear losses\label{LLBS}}

We consider wave packets $E=A\exp(-i\omega_0t+ik_0 z)$ oscillating a certain optical carrier frequency $\omega_0$ and of propagation
constant $k_0$, that self-focus symmetrically in all available dimensions. In two dimensions, this represents a monochromatic light beam $A(r,z)$
that depends only on the radial coordinate $r=(x^2+y^2)^{1/2}$ in the transversal plane. In three-dimensions, self-focusing is symmetric in a
medium with anomalous dispersion $[k_0^{\prime\prime}<0]$ if the envelope $A(r,z)$ depends only on the spatiotemporal radial coordinate
$r=(x^2+y^2+t^{\prime 2}/k_0|k_0^{\prime\prime}|)^{1/2}$, where $t'=t-k_0^{\prime}z$ is the local time and $k_0^{(n)}$ is the
nth-derivative of the propagation constant $k(\omega)$ at $\omega_0$. Two-dimensional symmetrical self-focusing in a planar wave can be
considered as well if $r=(x^2+t^{\prime 2}/k_0|k_0^{\prime\prime}|)^{1/2}$. In all these cases, the simplest model of self-focusing is the
nonlinear Schr\"odinger equation (NLSE)
\begin{equation}\label{NLSE}
\partial_z A =\frac{i}{2k_0}\Delta_r A+\frac{ik_0n_2}{n_0}|A|^2A
-\frac{\beta^{(M)}}{2}|A|^{2M-2}A,
\end{equation}
where $\Delta_r =\partial^2_r +[(D-1)/r]\partial_r$, with $D=2$ or $D=3$, and $n_0$ is the refractive index at $\omega_0$. A pure Kerr
nonlinearity with nonlinear refractive index $n_2>0$ is considered for simplicity, but other more complex self-focusing nonlinearities can be
considered as well. The term with $\beta^{(M)}>0$ accounts for nonlinear losses (NLLs) due to $M$-photon absorption.

The NLSE (\ref{NLSE}) supports localized and stationary solutions of the form $A=a(r)\exp[i\varphi(r)+ i\delta z]$, where $a$ and $\varphi$ are
real functions, with $\delta>0$ in completely transparent media (solitons), and with $\delta<0$ also in media with NLLs (linear or nonlinear
conical waves) \cite{DURNIN,JOHANNISSON,PORRAS1,PORRAS2}. Between these two families, Eq. (\ref{NLSE}) supports also localized and
stationary solutions without any axial wave vector shift ($\delta=0$) in nonlinear media with NLLs, that have been named lossy light bullets
\cite{PORRAS3,PORRAS5}, and that present substantially different characteristics from solitary or conical bullets. From Eq. (\ref{NLSE}), these
LLBs must satisfy
\begin{eqnarray}\label{ESTA}
a^{\prime\prime}+ \frac{D-1}{r} a' -\varphi^{\prime 2}a +\frac{2k_0^2n_2}{n_0}a^3&=&0 ,\\
\beta^{(M)}\!\!\!\int_0^{r}\!\!dr\pi(2r)^{D-1}a^{2M}&=&-\frac{\pi(2 r)^{D-1}}{k_0}\varphi' a^2,\label{ESTP}
\end{eqnarray}
where prime signs stand for $d/dr$, with boundary conditions with $a(0)=\sqrt{I_0}$, $a'(0)=0$, $\varphi'(0)=0$, and with the additional
localization condition $a\rightarrow 0$ for $r\rightarrow\infty$. These solutions are numerically to exist in two and three dimensions for all
peak intensity $I_0$ up to the maximum value
\begin{equation}\label{IMAX}
I_{0,\rm max}= \left[\frac{4\gamma_{\rm max} k_0 n_2}{n_0\beta^{(M)}}\right]^{\frac{1}{M-2}},
\end{equation}
where $\gamma_{\rm max}$ is a dimensionless parameter of the order of unity that depends slightly on $M$ and number of dimensions $D$.
Contrary to solitons, there no exist solutions to Eqs. (\ref{ESTA}) and (\ref{ESTP}) satisfying the localization condition in the case of only one
dimension.

\begin{figure}
\begin{center}
\includegraphics[height=3.8cm]{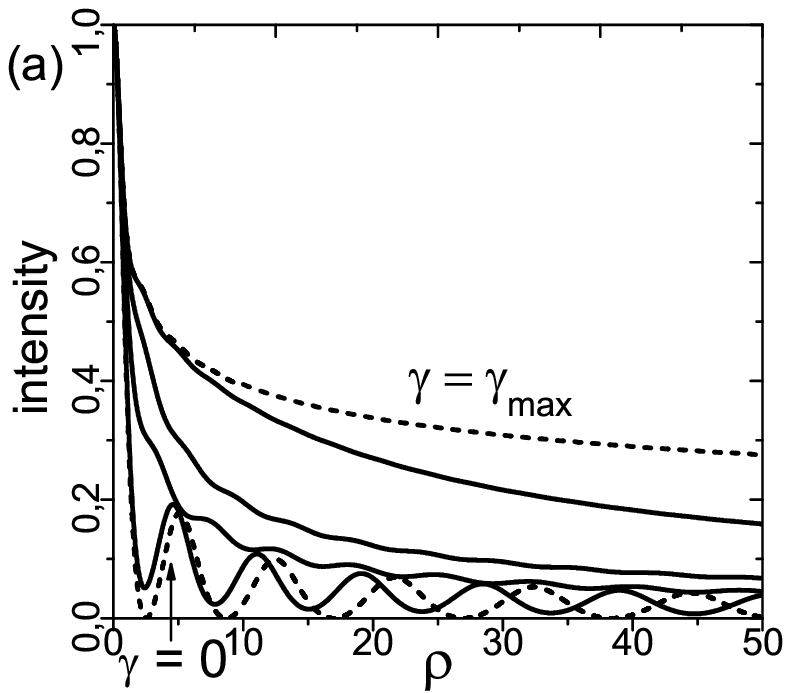} \includegraphics[height=3.6cm]{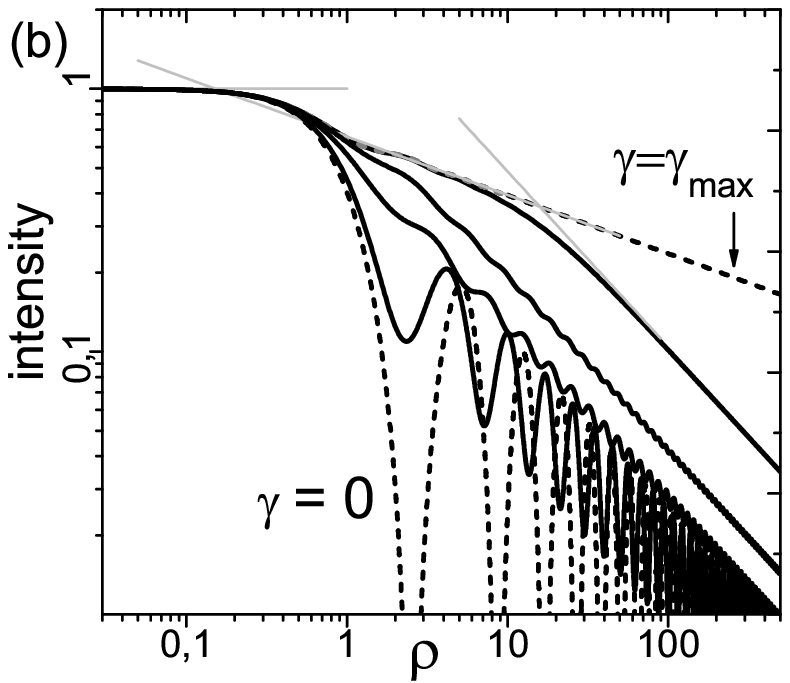}\\
\includegraphics[height=3.8cm]{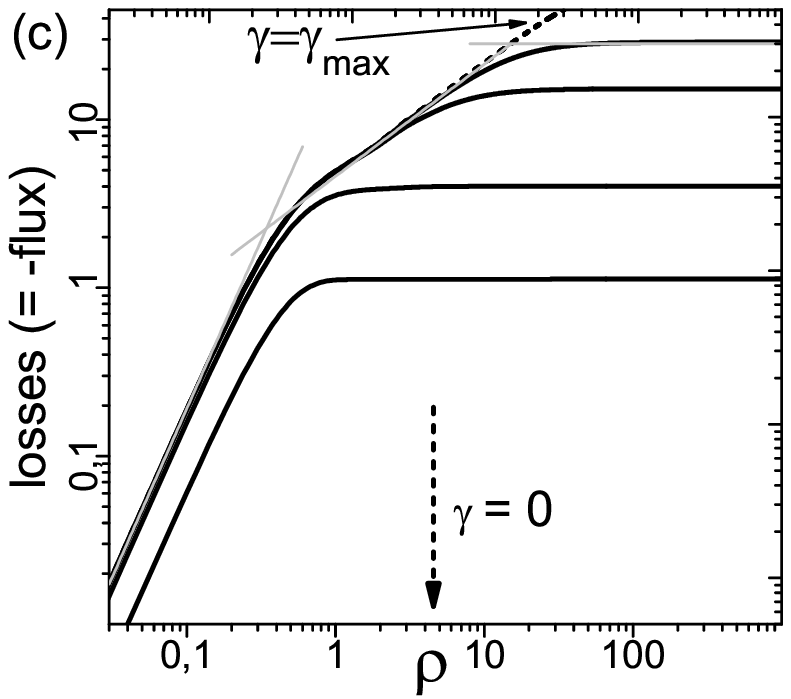} \includegraphics[height=3.8cm]{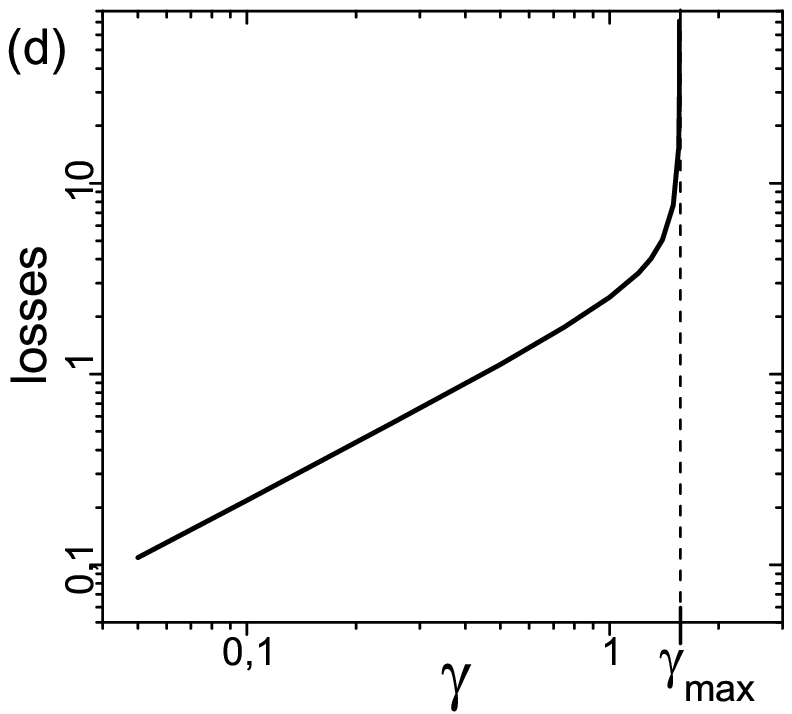}
\end{center}
\caption{(a) normalized intensity, (b) normalized intensity in double logarithmic scale, and (c) normalized NLLs (or normalized
inward power flux) radial profiles of LLBs with $M=6$ and $D=2$. The dashed curves correspond to the limiting cases of negligible and infinite
losses. They are obtained by solving (\ref{ESTA}) and (\ref{ESTP}) in the dimensionless form
$\tilde a''+(D-1)\tilde a'/\rho-\varphi^{\prime 2}\tilde a + 2 \tilde a^3=0$, and
$-\pi (2\rho)^{D-1}\varphi' \tilde a^2 = 4\gamma \int_0^{\rho}d\rho\pi(2\rho)^{D-1}\tilde a^{2M}$ (normalized inward flux $=$ normalized
losses), where $\tilde a=a/\sqrt{I_0}$, $\rho=\sqrt{k_0^2n_2I_0/n_0}\, r$ is the normalized radial coordinate, and with different values of
$\gamma=n_0\beta^{(M)}I_0^{M-2}/4k_0n_2=0.5,1.3,1.5,1.558$ between the limits $\gamma=0$ and $\gamma=\gamma_{\rm max}=1.5594$
for $M=6$ and $D=2$. (d) Normalized total NLLs, or normalized total inward flux, per unit propagation length as a function of
$\gamma$ (or peak intensity) up to the limiting value $\gamma_{\rm max}$ (or $I_{0,\rm max}$) of existence of LLBs.}
\label{fig1APB}
\end{figure}

Radial profiles for $D=2$ and $M=6$ with different values of $I_0$ between the limits $I_0\rightarrow 0$ of negligible losses (lower dashed curve)
and $I=I_{0,\rm max}$ (upper dashed curve) are shown in Figs. \ref{fig1APB}(a) and (b). Qualitatively similar profiles are found for different
values of $M$ and for $D=3$. LLBs are weakly localized so that they carry infinite power (in two dimensions) or energy (in three dimensions).
Equation (\ref{ESTP}), written as $N(r)=-F(r)$ for short, establishes that the nonlinearly absorbed power (energy) $N(r)$ per unit propagation
length in any circle (sphere) of radius $r$ must be refueled by identical inward radial power (energy) flux $-F(r)$ per unit length across the
circumference of the circle (surface of the sphere) for stationarity to be possible. This mechanism of stationarity with NLLs was first described for
conical light bullets ($\delta<0$) \cite{PORRAS1,PORRAS2}, but the existence of the LLBs evidences that it can also work without an
inward conical flux. The loss profiles $N(r)$ of the LLBs of Fig. \ref{fig1APB}(a) are represented in Fig. \ref{fig1APB}(c). LLBs loss a finite fraction
$N_T=N(r\rightarrow\infty)$ of their infinite power (energy) per unit propagation length, at the same time that a power (energy) flux $-F_T=N_T$
comes from large radial distances towards the center to compensate these losses. The value $-F_T=N_T$ is indicated as a horizontal gray straight
line in Fig. \ref{fig1APB}(c), and their values are represented in Fig. \ref{fig1APB}(d) for the different values of the peak intensity $I_0$. Only in
the limit $I_0=I_{0,\rm max}$, the localization of the lossy light bullet is so weak that both the power and the NLLs per unit propagation are
infinite.

\section{The structured profile of lossy light bullets and their energy reservoir \label{STRUCTURE}}

While in conical light bullets stationarity with NLLs is due to an overweight of the linear conical part pushing energy inward with respect to the
conical part pushing outward, in LLBs these contributions to the energy flux are not separable because of the nonlinear origin of the energy flux.
The dynamic balance of NLLs and self-focusing requires the structured radial profiles revealed in the double logarithmic plots of
Figs. \ref{fig1APB} (b) and (c), where three sequential scaling power-laws ($\propto r^{-\sigma}$) with different scaling powers $\sigma$ can
be appreciated  in each profile (gray straight lines).

First, while the intensity $a^2(r)$ remains of the order of $I_0$ in the vicinity of $r=0$ [horizontal straight line in Fig. \ref{fig1APB}(b)],
Eq. (\ref{ESTP}) yields $-F(r)=N(r)\simeq (2^{D-1}\pi r^D/D)\beta^{(K)}I_0^{(K)}$ [steepest straight line in (c)], and this flux requires wave
fronts $\varphi(r)\simeq \varphi(0)-k_0\beta^{(K)}I_0^{K-1}r^2/2D=\mbox{const.}$, i.e., spherical wave fronts with the same curvature
$1/R = \beta^{(K)}I_0^{K-1}/D$ at any propagation distance. About $r=0$, Eq. (\ref{ESTA}) reduces asymptotically to $a^{\prime\prime}
+(D-1)a^\prime/r+(2k_0^2n_2/n_0)a^3\simeq 0$. The shape of the central peak is then substantially independent of the dissipative properties
of the medium, and its width, $\Delta r\simeq [\ln (2) n_0 D /2 k_0^2n_2 I_0]^{1/2}$ (HWHM), is approximately equal to the width of the ground
solitons in two (Townes beam) and in three dimensions with the same peak intensity $I_0$ in transparent media. The central
peak of the LLB thus resembles a propagating soliton, but its is actually being continuously absorbed and replenished due to the flux
created by its converging wave fronts.

Second, replenishment of the whole central peak requires a larger flux coming from around [Figs. \ref{fig1APB}(b) and (c), intermediate straight
lines]. If we write the power law of the shell around the central peak as $a\simeq b r^{-\sigma}$,
where $b$ is a constant, Eq. (\ref{ESTP}) yields $-F(r)=N(r)\simeq c\, r^{D-2M\sigma}$, where
$c=\beta^{(M)}\pi 2^{D-1}b^{2M}/(D-2M\sigma)$. This requires converging (but non-spherical) wave fronts with
$\varphi'(r)\simeq -d\, r^{1-2(M-1)\sigma}$, where $d=k_0\beta^{(M)}b^{2M-2}/(D-2M\sigma)$.
From Eq. (\ref{ESTA}), the relation $-d^2r^{2-4\sigma(M-1)} + (2k_0^2n_2b^2/n_0)r^{-2\sigma}\simeq 0$ is found as the condition for the
strength of self-focusing to be consistent with these wave fronts. Equating the exponents and the constants of the two terms, one readily obtains
\begin{equation}\label{PL1}
\sigma_S=\frac{1}{2M-3},\quad b_S= \left[\sqrt{\frac{2n_2}{n_0}}\frac{(D-2M\sigma)}{\beta^{(M)}}\right]^{\frac{1}{2M-3}} ,
\end{equation}
for the power-law $a\simeq b_Sr^{-\sigma_S}$ of the amplitude in the shell surrounding the peak [intermediate straight line in
Figs. \ref{fig1APB}(b)]. This shell is wider, and the losses higher, as the peak intensity $I_0$ approaches $I_{0,\rm max}$, becoming infinitely
wide and high for the lossy light bullet with $I_0=I_{0,\rm max}$ [dashed lines in Figs. \ref{fig1APB}(b) and (c)].

Except for the lossy light bullet with $I_0=I_{0,\rm max}$, the outer part of the lossy light bullet is characterized by the absence of absorption,
since the inward flux reaches a constant value $-F_T=N_T$ that compensates the total NLLs $N_T$. Proceeding as above but replacing
$N(r)$ with $N_T$ at large $r$, one finds $\varphi'(r)\simeq - d\, r^{2\sigma-D+1}$, where $d=k_0 N_T/2^{D-1}\pi b^2$, for the converging wave
fronts, and the balance relation $-d^2r^{4\sigma -2D+2} + (2k_0^2n_2b^2/n_0)r^{-2\sigma}\simeq 0$
for the self-focusing state with these wave fronts. From this relation we find
\begin{equation}\label{PL2}
\sigma_a=\frac{D-1}{3},\quad b_a=\left(\frac{N_T^2n_0}{2^{2D-1}\pi^2 n_2}\right)^{1/6},
\end{equation}
for the asymptotic power-law $a\simeq b_ar^{-\sigma_a}$ at large $r$ [steepest straight line in Fig. \ref{fig1APB}(b)]. LLBs are then
nonlinear waves also asymptotically. Self-focusing is continuously pushing the power (energy) contained in this huge, widespread, and
non-absorbed reservoir to replenish the total lost power (energy) in the center.

In the context of the cubic-quintic Ginzburg-Landau equation, dissipative solitons with permanently converging wave fronts have also been
described in the two-dimensional case \cite{AKHMEDIEV,GRELU,ANKIEWICZ}. However, these dissipative solitons carry finite power, and their
stationarity is based on a balance between the losses and gain that take place in different parts of their radial profiles.

The radial oscillations accompanying the asymptotic decay of LLBs [Figs. \ref{fig1APB}(a) and (b)] have also little to do with the linear
oscillations of conical bullets. An asymptotic analysis similar to that described above shows that both the amplitude and frequency of the nonlinear
oscillations are proportional to the field amplitude $b_ar^{-\sigma_a}$. The amplitude and frequency of the oscillations follow then the same
scaling power law with power $(D-1)/3$, in contrast with the amplitude decay with power $(D-1)/2$ and the constant frequency of the oscillations
of conical light bullets. More precisely, at each radius $r$, the local radial frequency $K=(k^2+k_y^2)^{1/2}$, or spatiotemporal radial frequency
$K=[k_x^2+k_y^2+k_0|k_0^{\prime\prime}|(\omega-\omega_0)^2]^{1/2}$ of the oscillations about $b_ar^{-\sigma_a}$ are
given by
\begin{equation}\label{K}
K(r)\simeq k_0 \sqrt{\frac{12 n_2}{n_0}}b_ar^{-\sigma_a}.
\end{equation}
LLBs contain then a continuous of frequencies in its spectrum, with decreasing frequencies distributed along the radial profile
at increasingly larger distances from the bullet center.

\begin{figure}[t]
\begin{center}
\includegraphics[width=4.45cm]{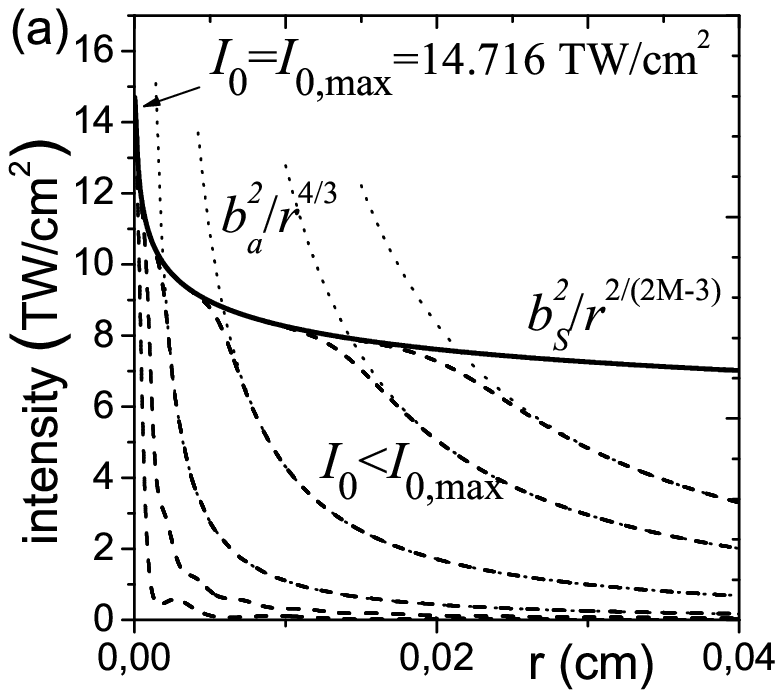}\includegraphics[width=4.35cm]{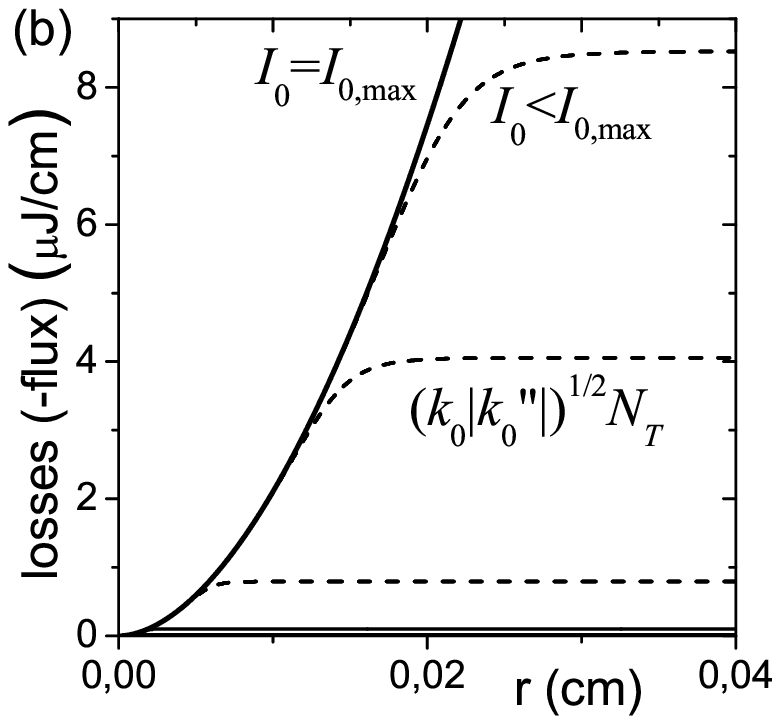}\\
\end{center}
\caption{\label{LBS} Solid thick curves: (a) Intensity and (b) nonlinear loss profiles $(k_0|k_0^{\prime\prime}|)^{1/2}N(r)$
[equal to the inward energy flux profile $-(k_0|k_0^{\prime\prime}|)^{1/2}F(r)$] of the most lossy three-dimensional light bullet, whose intensity
is $I_0=I_{0,\rm max}=14.716$ TW/cm$^2$ [Eq. (\ref{IMAX}) with $\gamma_{\rm max}=3.15294$ for $D=3$ and $M=10$] in fused silica at
1550 nm carrier wave length ($k_0=5.854\times 10^4$ cm$^{-1}$, $k_0^{\prime\prime}=-279.4$ cm$^{-1}$fs$^2$, $n_2=2.2\times 10^{-16}$
cm$^2$/W, $\beta^{(M)}=5.11\times 10^{-116}$ cm$^{17}/$W$^9$, and $M=10$). Dashed thick curves: Intensity and nonlinear loss profiles of
LLBs with slightly lower peak intensities $I_0$, for comparison.}
\end{figure}

\subsection{The most lossy light bullet in a nonlinear dissipative medium \label{MLLB}}

Oscillations cease in the limit of peak intensity $I_0$ equal to $I_{0,\rm max}$. As we will see in Sec. \ref{ATTRACTIVE}, this limiting bullet with
infinite power and infinite losses is, despite its ideal nature, of particular relevance in the self-focusing of real (finite-energy)
wave packets in media with NLLs. Its peak intensity, given by Eq. (\ref{IMAX}), depends only on the number of dimensions and the
optical properties of the medium at the carrier wave length. The lossy shell extends up to infinite radial distances, and therefore its asymptotic
decay follows the power law specified by Eq. (\ref{PL1}), which depends only on the number of dimensions and the optical properties of the
medium.

For reference in Section \ref{ATTRACTIVE}, the solid curves in Figs. \ref{LBS}(a) and (b) represent the radial profiles of intensity and NLLs
(-flux) of the most lossy, three-dimensional light bullet in fused silica at 1550 nm carrier wavelength ($M=10$). The intensity asymptotic decay
$b_S^2/r^{2/(2M-3)}=b_S^2/r^{2/17}$ is much slower than the intensity asymptotic decays $b_a^2/r^{2(D-1)/3}=b_a^2/r^{4/3}$ (dotted
curves) of the LLBs of lower intensity, shown in Fig. \ref{LBS}(a) as dashed curves.

\section{Propagation properties of physically realizable lossy light bullets \label{FINITE}}

Since LLBs carry infinite energy, only approximate versions with finite energy can be formed in practice. As seen in Sec. \ref{STABILITY}, LLBs
with $I_0$ close to $I_{0,\rm max}$ are spontaneously formed, but only up to a maximum radial distance $r_t$, from the finite-energy of
collapsing Gaussian-like wave packets in media with NLLs.

\begin{figure*}[t]
\begin{center}
\includegraphics[width=5cm]{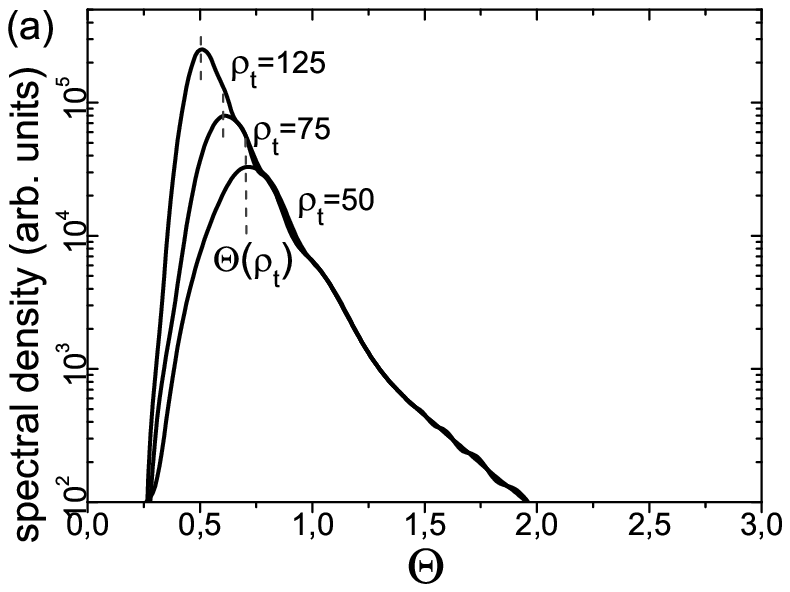}
\includegraphics[width=5cm]{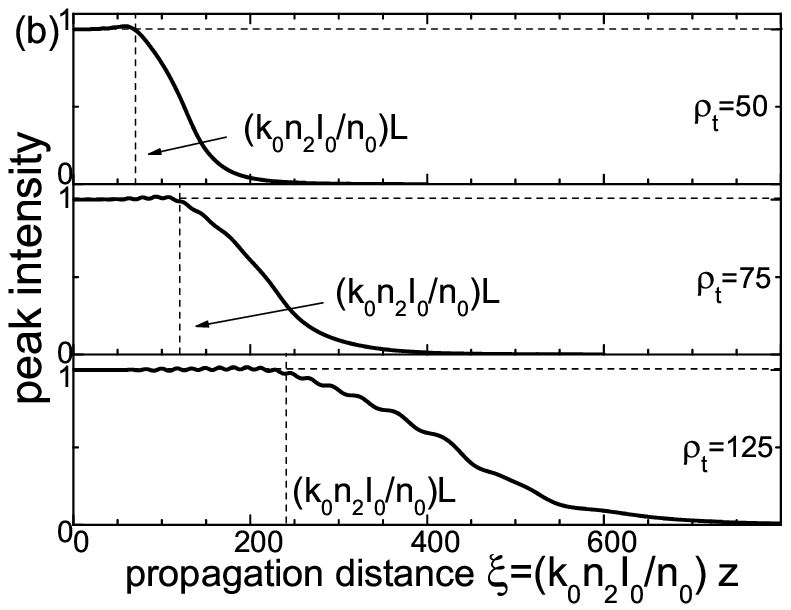}
\end{center}
\caption{(a) Radial spectral intensity of two-dimensional LLBs $\tilde a\exp(i\varphi)$ with $D=2$, $M=6$ and $\gamma=1.3$, truncated at
different radii $\rho_t$. The truncation function is the super-Gaussian function $\exp(-\rho^4/\rho_t^4)$. The maxima are located at the local
frequency $K(\rho_t)$ at the truncation radius $\rho_t$ (vertical segments) as given by Eq. \ref{K}. The normalized radial frequency in the graphic
is $\Theta=K/\sqrt{k_0^2n_2I_0/n_0}$. (b) Axial evolution of the peak intensity for the truncated LLBs in (a). It has been obtained by solving the
NLSE (\ref{NLSE}) in the dimensionless form $\partial_\xi\tilde A=(i/2)\Delta_\rho\tilde A+i|\tilde A|^2\tilde A-
2\gamma|\tilde A|^{2M-2}\tilde A$ taking truncated LLBs $\tilde A(\rho,0)=\tilde a\exp(i\varphi) \exp(-\rho^4/\rho_t^4)$ as initial
conditions. The normalized amplitude $\tilde a$, phase $\varphi$, radius $\rho$, and $\gamma$ are defined in Fig. \ref{fig1APB}. The
normalized axial coordinate is $\xi=(k_0n_2I_0/n_0) z$. The dashed vertical segments indicate the light bullet propagation distances $L$ for each
truncation $\rho_t$ predicted by Eq. (\ref{LBD}).}
\label{fig2}
\end{figure*}

The propagation properties of these truncated LLBs can be understood from the structure described above. For example, Fig.
\ref{fig2}(a) shows the spectrum of radial frequencies of two-dimensional LLBs truncated at different radii $r_t$. Truncation removes
the frequencies $K<K(r_t)$ distributed at $r>r_t$, as in a low-pass filter. This causes the frequencies $K\simeq K(r_t)$ distributed about the
truncation radius $r_t$ to dominate, since they fill the largest area or volume. The spectrum then resembles the annular spectrum of a conical
wave with a main cone angle $K(r_t)/k_0$ determined by the truncation radius. This may be the reason why LLBs with finite radius
can easily interpreted as (linear or nonlinear) Bessel beams in self-focusing experiments \cite{FACCIO}. The frequency spectrum of truncated
LLBs differ from those of truncated conical bullets in that higher frequencies, $K>K(r_t)$, placed at $r< r_t$ have more weight
than in conical waves. Indeed, a truncated lossy light bullet behaves very much as a continuous of conical light bullets with cone
angles equal or larger than $K(r_t)/k_0$, as seen below.

Despite the central peak of the lossy light bullet is continuously being absorbed, it can propagate without significant change for hundred times the
diffraction length $k_0(\Delta r)^2/2$ associated to its width (in the radially symmetric two-dimensional case), or hundred times the dispersion
length, equal to the diffraction length, associated to its duration (in the spatiotemporal radially symmetric three-dimensional case). The distance of
light bullet behavior $L$ is determined by the distance at which the energy reservoir of the truncated lossy light bullet is consumed. For example,
taking the truncated LLBs with the spectra of Fig. \ref{fig2}(a) as initial conditions, the evolution of the peak intensity along $z$ according to the
NLSE (\ref{NLSE}) is plotted in Fig. \ref{fig2}(b), where the normalized axial coordinate in Fig. \ref{fig2}(b) is such that the diffraction length
associated to the width of the central peak is unity.

To estimate the distance $L$ we consider the truncated lossy light bullet as a superposition of conical waves with cone angles
$\theta \simeq K(r)/k_0 \ge K(r_t)/k_0$. Thus, the power coming from a position $r<r_t$ in the initial condition reaches the center of the bullet,
refilling it, at the diffraction-free distance $r/\theta = k_0 r/K(r)$ \cite{DURNIN}, or from Eq. (\ref{K}), at $k_0r^{(D+2)/3}/\sqrt{12}b_a$.
In particular, the power coming from the bullet border $r_t$ reaches the center at the longest distance
\begin{equation}\label{LBD}
L = \frac{k_0}{\sqrt{12}\,b_a}r_t^{\frac{D+2}{3}},
\end{equation}
which estimates the distance at which the power reservoir is completely consumed, and therefore the central peak initiates to decay. As seen in
Fig. \ref{fig2}(b), Eq. (\ref{LBD}) gives indeed the distance at which the peak intensity initiates to disappear. Unlike conical waves, the light bullet
distance $L$ grows faster than linearly with truncation radius, and depends also on the parameter $b_a$ related to the total NLLs
$N_T$ per unit propagation length of the specific LLB.

\section{Self-reconstruction property  \label{SELF}}

\begin{figure}[b]
\begin{center}
\includegraphics[width=4.2cm]{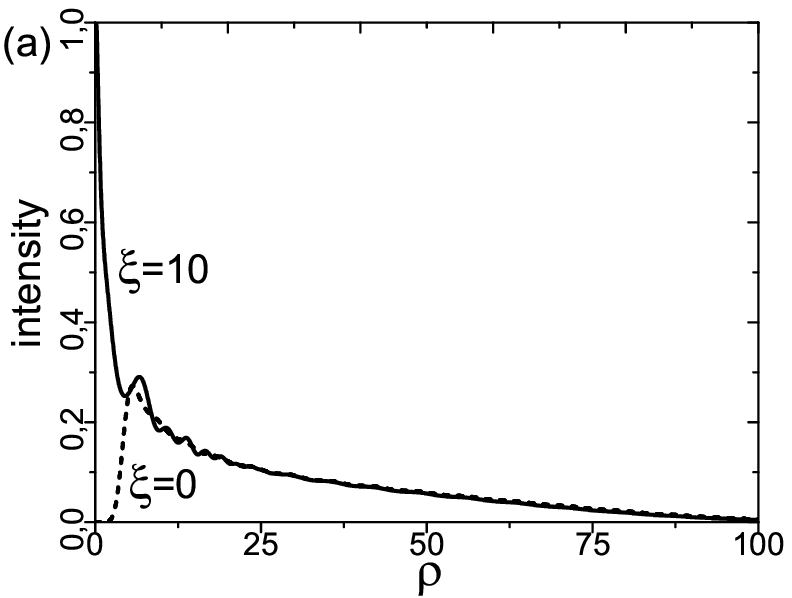} \includegraphics[width=4.2cm]{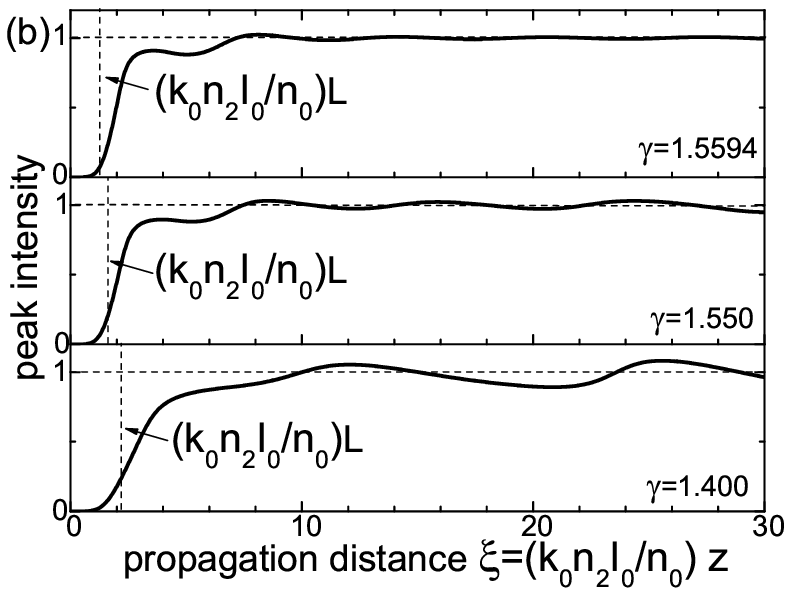}\\
\end{center}
\caption{(a) LLB $\tilde a\exp(i\varphi)$ with $D=2$, $M=6$ and $\gamma=1.5$ (dashed curve) blocked up to $\rho_t=4$ with block
function $1-\exp(-\rho^4/\rho_t^4)$, and its propagated field (solid curve). (b) Change of the peak intensity
with propagation distance for the LLBs with $D=2$, $M=6$ and $\gamma=1.5594$, $1.55$, $1.4$, all blocked up to the same radius
$\rho_t=4$. Normalized field, propagation equation, coordinates and parameters are as in Figs. (\ref{fig1APB}) and (\ref{fig2}).
The dashed vertical lines indicate the estimate of self-reconstruction distances $L$ given by Eq. (\ref{LBD}).}
\label{fig3}
\end{figure}

As conical bullets \cite{BOUCHAL}, LLBs have the property of rebuilding after being partially blocked. The dashed curve in Fig. \ref{fig3}(a)
represents a two-dimensional lossy LLB with a hole of radius $r_t$ comparable to that of the central peak, and the solid curve the propagated field
from the NLSE (\ref{NLSE}) at a distance where the central peak is reconstructed. The self-reconstruction distance can be estimated following a
similar reasoning as above. The peak initiates to rebuild when the power surrounding the hole begins to reach the center. For the different cone
angles $K(r)/k_0$ at different radii $r\ge r_t$, the shortest axial distance of arrival of power corresponds to $r_t$, and is then given again by
Eq. (\ref{LBD}), with $r_t$ standing now for the radius of the hole. For several two-dimensional LLBs with different peak intensities $I_0$ and
blocked up to a given radius $r_t$, Fig. \ref{fig3}(b) evidences that the peak intensity rises from zero up to the corresponding
unblocked values $I_0$ at a distance that is well approached by Eq. (\ref{LBD}). Being equal the hole radii, different values of $L$ are due only to
different values of $b_a$, i.e., the different total NLLs $N_T$ of the input LLBs. Of course, these numerical simulations have been performed in a
grid of finite radius, and therefore the self-reconstruction property holds for LLBs with finite power.

\section{Stability properties  \label{STABILITY}}

Since LLBs are nonlinear waves, they can suffer from stabilities when they are perturbed, e. g., by truncation at $r>r_t$ or by obscuration at
$r<r_t$, as in the preceding sections. In fact, rebuilding of the central peak is seen in Fig. \ref{fig3}(b) to be accompanied by nonlinear
oscillations, but these oscillations disappear as the intensity $I_0$ of the rebuilding lossy light bullet approaches $I_{0,\rm max}$. Also, the
truncated LLBs of Fig. \ref{fig2}(b) develop nonlinear oscillations, but again their growth rate is smaller as $I_0$ approaches the limit
$I_{0,\rm max}$. In this section we show that NLLs confer stability to these light bullets. The LLBs of the maximum intensity and losses
is consequently the most stable among all LLBs, which acts as an attractor in the self-focusing dynamics with nonlinear losses.

From numerical simulations, we first show that the most lossy light bullet tends to be spontaneously formed in the collapse of standard,
Gaussian-like wave packets arrested by NLLs. Of course, the complete formation of the this LLB attractor would require infinite amounts of energy.
It is then only formed up to a certain maximum radius, depending on the available energy in the initial Gaussian wave packet, and its incomplete
formation causes its eventually decay, which takes place through an adiabatic sequence of less dissipative LLBs. For conciseness, we review these
results in the case of three-dimensional, or spatiotemporal collapse of ultrashort pulses in media with anomalous dispersion \cite{PORRAS3}.
Similar results for the two-dimensional, spatial collapse of monochromatic light were described in \cite{PORRAS4}.

The dynamics of the spontaneous formation and decay of the most lossy light bullet described by the simple Kerr$+$NLL model is found to
describe the observed facts in the dynamics of light filaments excited by ultrashort pulses in media with anomalous dispersion
\cite{MOLL,SKUPIN}, including the filament intensity, the formation of long filament segments and repeated collapse in the form of short bursts,
which supports the relevance the most lossy light bullet in these filamentation experiments.  In this connection, the most lossy three-dimensional
light bullet appears here as the counterpart in media with anomalous dispersion of the conical light bullets that are spontaneously formed in
media with normal dispersion \cite{TRAPANIX,FACCIOPRL,KOLESIK1}, and would constitute an alternate form the so-called light bullets
\cite{SILBERBERG,TRAPANI,EISENBERG,FIBICH}, or stable, stationary and localized wave packets in all dimensions.

Second, we perform a linearized analysis of instability of LLBs that shows that they are unstable under radial perturbations. The exponential gain
of the unstable modes, however, decrease drastically with increasing NLLs, which confirms the stabilizing role of NLLs. In particular, the gain of the
most lossy light bullet is found to be negligible or zero, which is in agreement with its attractive property.

\subsection{The most lossy light bullet as an attractor of the self-focusing dynamics with nonlinear losses \label{ATTRACTIVE}}

\begin{figure}[!tbp]
\begin{center}
\includegraphics[width=8.5cm]{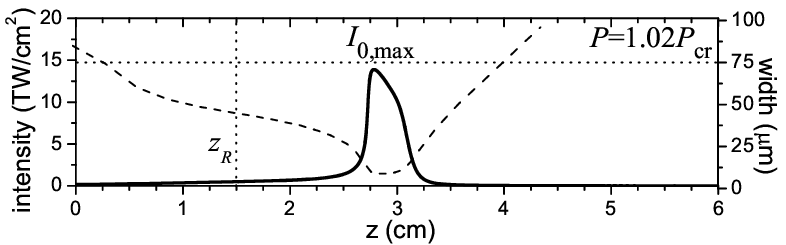}\\
\includegraphics[width=8.5cm]{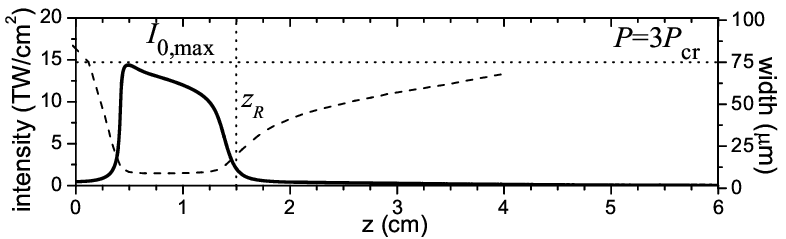}\\
\includegraphics[width=8.5cm]{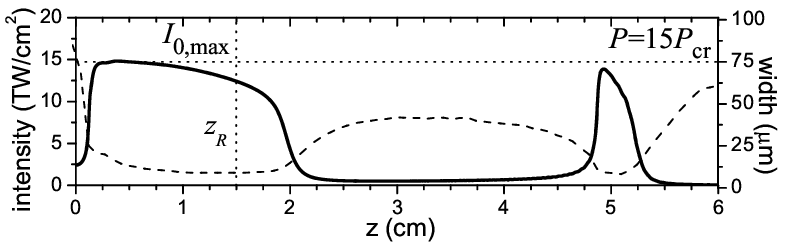}
\end{center}
\caption{\label{AXIAL} Peak intensity (solid curves) and FWHM width (dashed blue curves) along
$z$ for input Gaussian pulses of carrier frequency $\omega_0=1.21525$ fs$^{-1}$,
Gaussian width $s=0.00716$ cm [duration
$s(k_0|k_0^{\prime\prime}|)^{1/2}=29$ fs], and increasing power above the critical
power $P_{\rm cr}=13$ MW, calculated from Eq. (\ref{NLSE}) with the parameters of
fused silica ($k_0=5.854\times 10^4$ cm$^{-1}$, $k_0^{\prime\prime}=-279.4$
cm$^{-1}$fs$^2$, $n_2=2.2\times 10^{-16}$ cm$^2/$W, $\beta^{(M)}=5.11\times 10^{-116}$
cm$^{17}/$W$^9$, with $M=10$ \cite{SKUPIN}).}
\end{figure}

Figure \ref{AXIAL} shows the axial evolution of the peak intensity and width when the three-dimensional (spatiotemporal symmetric) Gaussian
pulses
\begin{equation}
A=\sqrt{\frac{2P}{\pi s^2}}\exp\left(-\frac{r^2}{s^2}\right),
\end{equation}
of carrier wavelength $\lambda_0=2\pi c/\omega_0=1550$ nm, and of different
peak powers $P$ above the critical peak power for spatiotemporal self-focusing \cite{NOTA},
\begin{equation}
P_{\rm cr}= \frac{2.157\lambda_0^2}{4\pi n_0n_2},
\end{equation}
are launched in fused silica. The peak intensity and width are evaluated from the numerical solution of the NLSE (\ref{NLSE}) including only the
Kerr and loss nonlinearities.  Increasing the energy of the input pulse, the light ``segments" of nearly constant high intensity and narrow width
become longer, and a number of light ``bursts"
beyond the segment are formed. These segments may extend beyond the Rayleigh distance $z_R=\pi s^2/\lambda_0$ of the input pulse,
and survive by several hundred times the Rayleigh distance expected from their width. These facts have been observed in self-channelling
experiments and numerical simulations \cite{MOLL,SKUPIN}. The location and intensity of the segments and bursts in Fig. \ref{AXIAL} are even in
quantitative agreement with accurate simulations under the same conditions (Fig. 2 in \cite{SKUPIN}) that include in the NLSE not only Kerr
nonlinearity and NLLs, but also all relevant higher-order effects in propagation (space-time focusing, self-steepening, higher-order dispersion and
plasma defocusing), and using input Gaussian pulses that are not completely symmetric.

\begin{figure}
\begin{center}
$\,\,\,$\includegraphics[width=8.5cm]{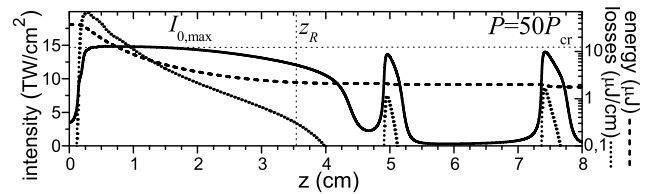}
\end{center}
\caption{\label{LAST} Peak intensity (solid curve), energy (dashed curve) and losses
(dotted curve) along $z$ in the same conditions as in Fig. \ref{AXIAL} except the
Gaussian width $s=0.011$ cm [Gaussian duration
$s(k_0|k_0^{\prime\prime}|)^{1/2}=44.5$ fs] and the peak power $P=50P_{\rm cr}$.}
\end{figure}

Remarkably, the peak intensity of the segments and bursts reaches a value that is close to the intensity of the LLB with the maximum intensity
$I_{0,\rm max}$ and infinite losses. The intensity of the segments and burst does not depend on the initial power, as in Figs. \ref{AXIAL}
from (a) to (c), and of its width, as seen in Fig. \ref{LAST}). Also, in experiments and ``exact" numerical simulations of self-focusing in fused
silica at the same carrier wave length, the intensity of the long-lived quasi light bullets, or filament segments, is close to the intensity
$I_{0,\rm max}$ of the most lossy light bullet. Along the segments, however, the pulse energy is drastically decreasing [Fig. \ref{LAST}, dashed
curve] due to an energy loss per unit length [Fig. \ref{LAST}, dotted curve] comparable to the total energy. These
observations suggest that collapse arrested by NLLs results in the (partial) formation of the LLB of the intensity
$I_{0,\rm max}$. Among all possible LLBs with different intensities, the equilibrium between self-focusing and NLLs
appears to be more stably reached in the LLB with maximum losses.

\begin{figure*}
\begin{center}
\includegraphics[width=3.8cm]{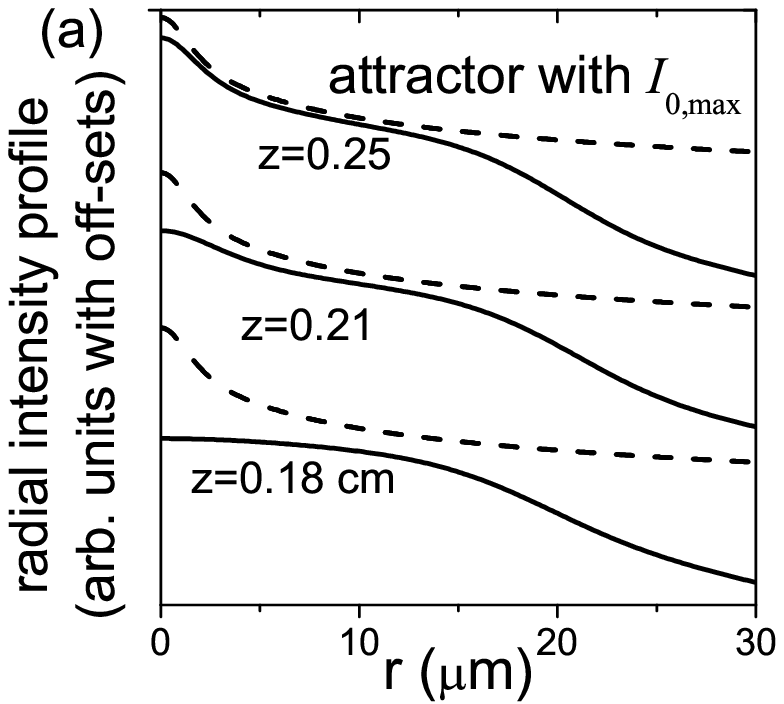}\includegraphics[width=3.9cm]{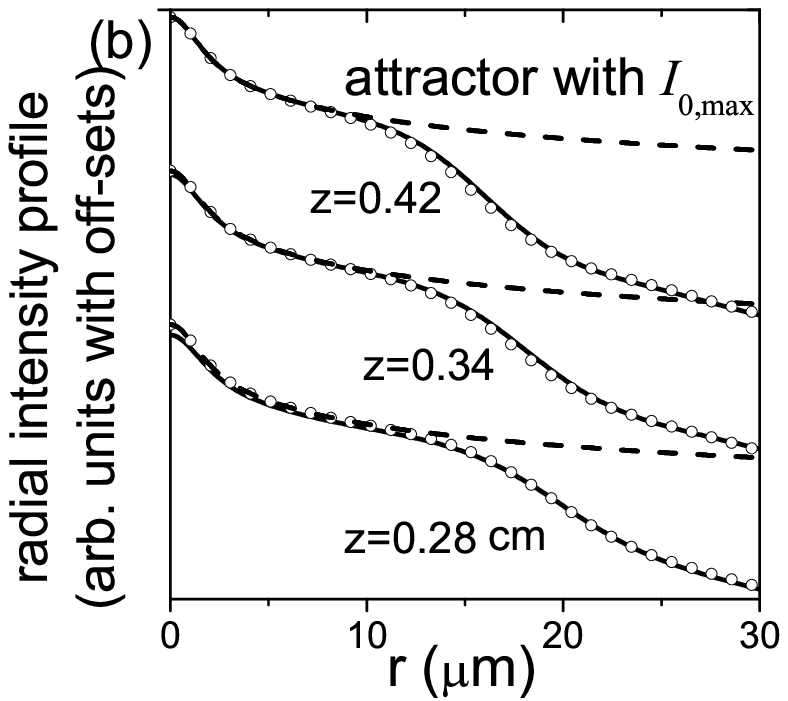}
\includegraphics[width=3.8cm]{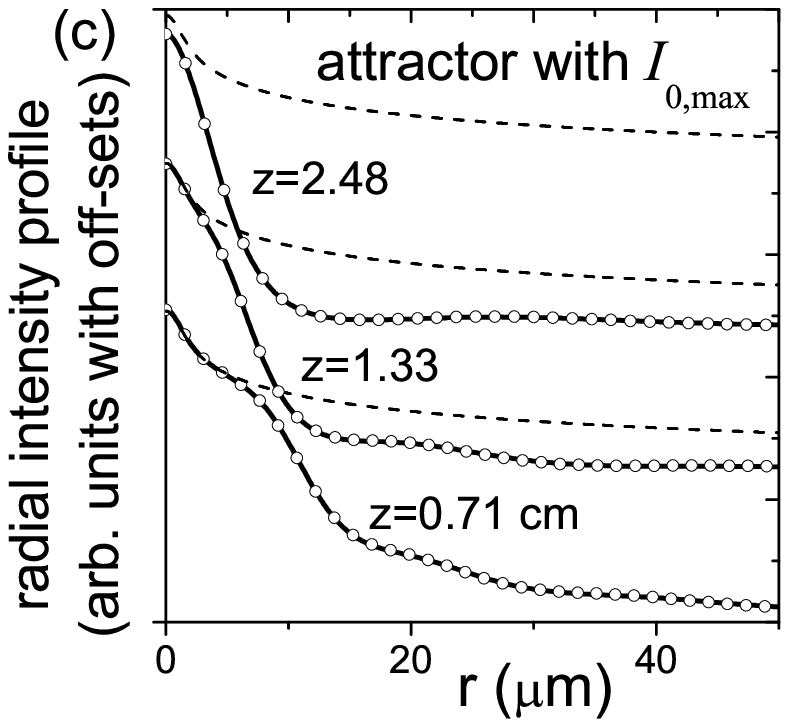}
\end{center}
\caption{\label{LAST2} For the same simulation as in Fig. \ref{LAST},
radial intensity profiles of the pulse at increasing propagation distances $z$
(solid curves), of the most lossy light bullet (dashed curve) and of the lossy or conical light bullet fitting the
pulse (open circles). A vertical off-set is introduced for clarity.}
\end{figure*}

Also, the radial intensity profile of the input Gaussian wave packet transforms during the self-focusing stage into the radial profile of most
LLB [Fig. \ref{LAST2} (a), solid curves and dashed curves]. Once the intensity is stabilized at about $I_{0,m}$, the inner part of the radial
profile fits accurately that of the most LLB [Fig. \ref{LAST2}(b), solid and dashed curves]. However, the finite energy of the pulse prevents the
pulse from reaching completely this bullet attractor. As $z$ increases, the slowly evolving radial profile along the segment fits successively the
profiles of LLBs with slowly changing intensities extremely close to $I_{0,m}$ [Fig. \ref{LAST2}(b), open circles]. At longer propagation distances,
the increasingly lack of energy forces the pulse to decay into less dissipative light bullets, endowed momentarily of a very small cone angle. The
radial profiles [Fig. \ref{LAST2}(c), solid curves] at different values of $z$ are seen to match now the those of conical light bullets with the same
peak intensity and total losses as the propagating pulse [Fig. \ref{LAST2}(c), open circles]. Contrary to what is stated in Ref. \cite{PORRAS4} in
the two-dimensional case, the cone angles in the relaxation stage are small but not completely negligible.

\begin{figure}[!tbp]
\begin{center}
\includegraphics[width=5.5cm]{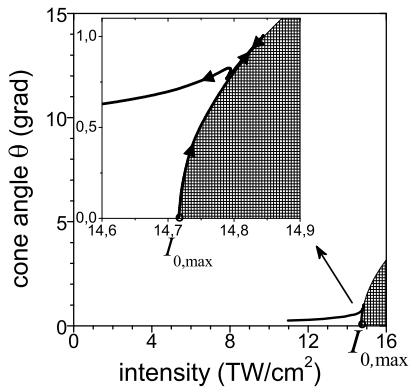}
\end{center}
\caption{\label{LLAST} For the same
example as in Figs. \ref{LAST} and \ref{LAST2}, visualization of the evolution as a trajectory in the parameter space of
conical and lossy light bullets in fused silica at $1550$ nm. The inset shows a tiny region close to $I_{0,\rm max}$.}
\end{figure}

Figure \ref{LLAST} offers an overall view of the collapse, filamentation, and relaxation dynamics in the space of parameters of conical light bullets
($I_0,\theta$), and in the subspace of LLBs ($I_0,\theta=0$). The dashed area indicates the region of parameters where these bullets
do not exist. Self-focusing carries the pulse directly to the point ($I_{0,\rm max},0$) representing the LLB of maximum losses in fused silica at
1550 wave length, or to points so close to it that they cannot be discerned at the scale of the figure or of the inset, remaining in this vicinity for
about the first half of the collapse segment. Relaxation follows the indicated trajectory, where it is seen that the cone angle grows initially, but
returns to zero at the end of the segment. The same dynamics explains also the ``bursts" after the segments, if any, but since the remaining
energy is considerably smaller, the lossy light bullet attractor is less approached and relaxation is faster.

\subsection{Stability under small perturbations \label{ANALYSIS}}

The attracting property of the most LLB suggests that it must be endowed of certain stability properties. An analysis of stability of
these LLBs, as done for other solitons or nonlinear conical waves, turns out to be an insurmountable task because, first, they are known only
numerically, and second, and mainly, because they are very weakly localized. We follow instead a simplified procedure, close to that in
\cite{SOTO}, to show that LLBs become more stable as their intensity and losses increase,  the most lossy light bullet being therefore the most
stable LLB. The idea that NLLs confer stability has already been expressed in \cite{PORRAS1} and \cite{ARNAUD} in relation to conical
beams in media with NLLs. In absence of a conical structure, the spontaneous transformation of Gaussian-like wave packets into the
most LLB upon self-focusing reflects the fact that this LLB is the most stable, non-conical, stationary state
in a self-focusing medium with NLLs.

To simplify the analysis we use the dimensionless variables $\tilde A=A/\sqrt{I_0}$, $\rho=\sqrt{k_0^2n_2I_0/n_0}\,r$, and
$\xi=(k_0n_2I_0/n_0)z$, to rewrite the NLSE (\ref{NLSE}) as
\begin{equation}\label{NLSE2}
\partial_\xi \tilde A =\frac{i}{2}\Delta_{\rho} \tilde A+i|\tilde A|^2\tilde A
-2\gamma |\tilde A|^{2M-2}\tilde A ,
\end{equation}
where $\Delta_{\rho} =\partial_\rho^2 +[(D-1)/\rho]\partial_\rho$, and
\begin{equation}
\gamma=\frac{n_0\beta^{(M)}I_0^{M-2}}{4 k_0 n_2} .
\end{equation}
The equations for LLBs $\tilde A=\tilde a \exp[i\varphi]$ become
\begin{eqnarray}
\label{ESTA2}
\tilde a^{\prime\prime}+ \frac{D-1}{\rho} \tilde a' -(\varphi^{\prime})^2\tilde a +2 \tilde a^3&=&0 ,\\
4\gamma \int_0^\rho d\rho \pi (2\rho)^{D-1}\tilde a^{2M}&=& -\pi (2\rho)^{D-1}\varphi' \tilde a^2 , \label{ESTP2}
\end{eqnarray}
where the last equation admits also the differential form
\begin{equation}
\left(\varphi^{\prime\prime}+ \frac{D-1}{\rho} \varphi' \right)\tilde a^2 +\varphi'(\tilde a^2)^\prime +4 \gamma \tilde a^{2M}=0 ,
\label{ESTP22}
\end{equation}
with boundary conditions $\tilde a'(0)=0$, $\varphi'(0)=0$ and $\tilde a(0)=1$, and the localization condition $\tilde a\rightarrow 0$ for
$\rho\rightarrow\infty$. The parameter
$\gamma$ ranges from 0 for the lossless case to the value of $\gamma_{\rm max}$ of the limiting LLB for the given values of $D$ and $M$.

In a standard, linearized analysis of stability, a weakly perturbed steady state of the form
\begin{equation}\label{PERT}
\tilde A(\rho,\xi) = \tilde a(\rho)e^{i\varphi(\rho)} + \epsilon\left[u(\rho)e^{i\kappa\xi}+v^\star(\rho)e^{-i\kappa^\star\xi}\right]
\end{equation}
is introduced into Eq. (\ref{NLSE2}), and nonlinear terms in the small parameter $\epsilon$ are disregarded. This leads to the differential
eigenvalue problem
\begin{equation}\label{EIGEN}
\left(\begin{array}{cc} H & f \\ -f^\star & -H^\star \end{array}\right)
\left(\begin{array}{c} u \\ v \end{array}\right) =
\kappa \left(\begin{array}{c} u \\ v \end{array}\right),
\end{equation}
where $H=\frac{1}{2}\Delta + (2\tilde a^2 +2i\gamma M\tilde a^{2M-2})$ and $f=[\tilde a^2 + 2i\gamma(M-1)\tilde a^{2M-2}]e^{2i\varphi}$.
Stability is determined by the absence of eigenvalues $\kappa=\kappa_R+i\kappa_I$ with negative imaginary part $\kappa_I$, which would lead
to an exponential growth with gain $-\kappa_I$ of the associated eigenmode $(u,v)$. Similar eigenvalue problems have been solved numerically
in the case of solitons in lossless media. For nonlinear Bessel beams, a numerical solution becomes barely practicable. Due to their persistent
tails, a huge radial grid with thousand of points is needed, and truncation, even if weak, tends to falsify the spectrum of eigenvalues.
Partial results for the two-dimensional case were nevertheless obtained in Refs. \cite{PORRAS1,ARNAUD} in relation to the stabilizing role of
losses in these nonlinear Bessel beams. LLBs are even less localized than nonlinear Bessel beams, and these difficulties become overwhelming.

Instead of solving (\ref{EIGEN}), we have followed the simplified procedure of launching weakly perturbed LLBs as initial conditions
and letting the possible exponential instability to be manifested. If the initial perturbation is weak enough, all excited unstable modes will growth
from very low values, the unstable mode with the highest exponential rate $-\kappa_I$ will emerge from others at a distance where it is still a
small (linear) perturbation, and its eigenvalue $\kappa$ and shape $(u,v)$ can be easily extracted.

\begin{figure}[h]
\begin{center}
\includegraphics[width=7.0cm]{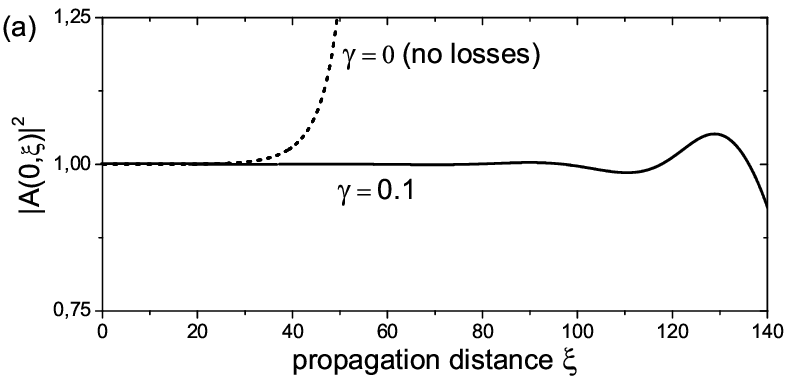}
\hspace*{0.35cm}\includegraphics[width=6.8cm]{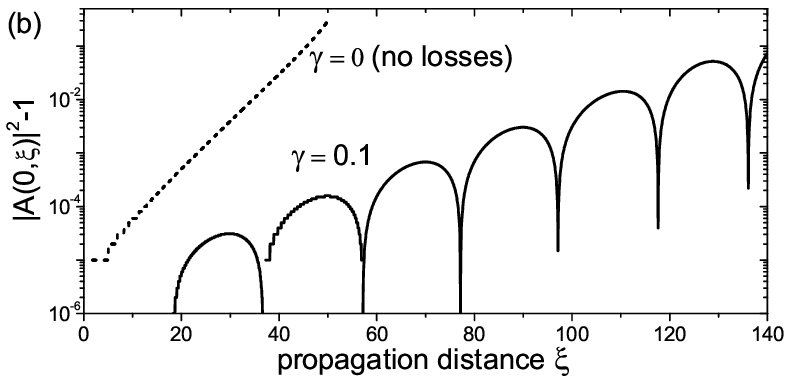}
\caption{\label{instability1} (a) Normalized axial intensity $|A(0,\xi)|^2$ and (b) difference $|A(0,\xi)|^2-1$
of initially perturbed (truncated at the large radius $\rho_t=1132$) three-dimensional LLBs in
media with $M=0$ and intensities such that $\gamma\simeq 0$ (dashed curves) and $\gamma=0.1$ (solid curves).}
\end{center}
\end{figure}

\begin{figure}[b]
\begin{center}
\includegraphics[width=4cm]{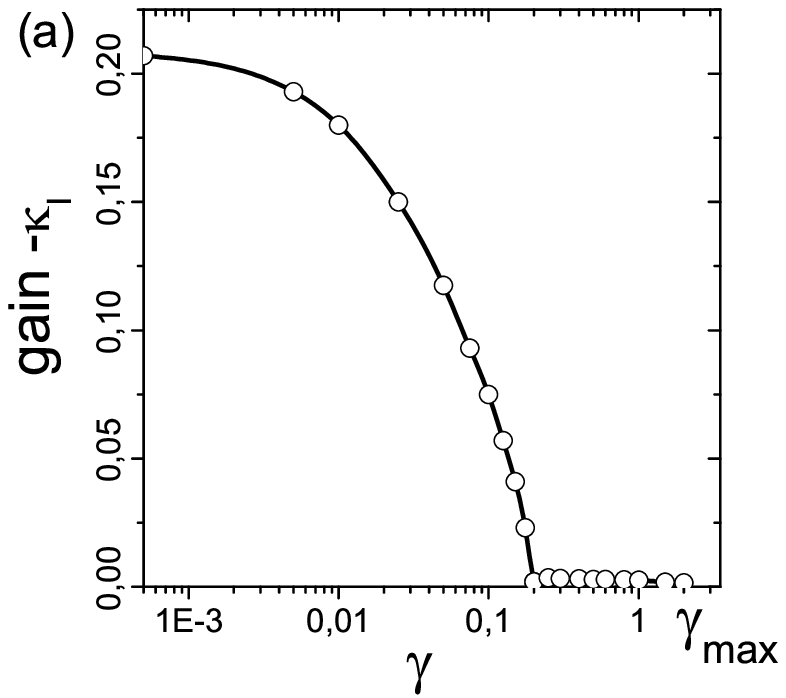}\includegraphics[width=4.1cm]{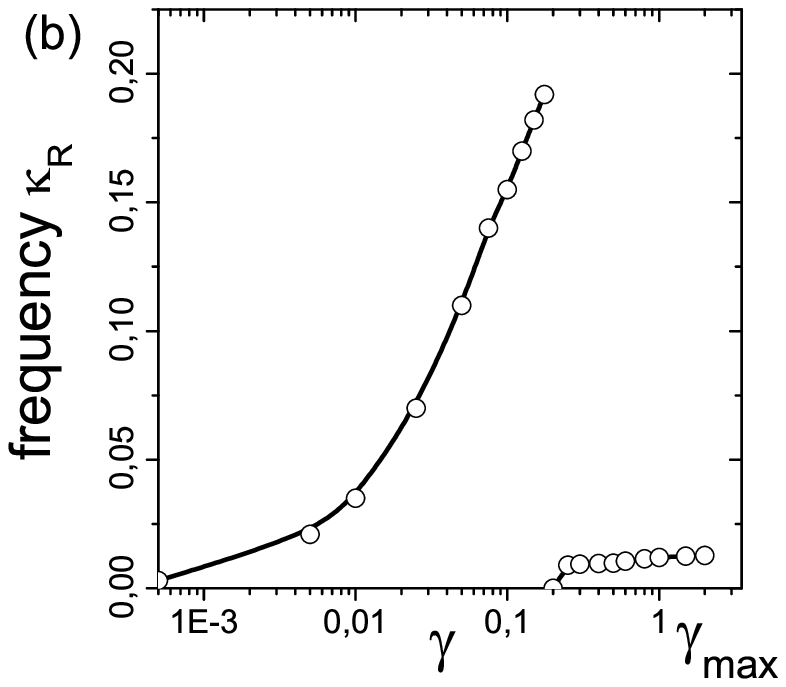}
\caption{\label{instability2} Normalized exponential gain $-\kappa_I$ and axial oscillation frequency $\kappa_R$ of the dominant unstable mode
of three-dimensional LLBs with $M=10$ as functions of increasing losses $\gamma$. In physical units the gain and frequency are
$-(k_0n_2I_0/N_0)\kappa_I$ and $(k_0n_2I_0/N_0)\kappa_R$, respectively.}
\end{center}
\end{figure}

Truncation of the LLB at a large radius is used as the weak perturbation that onsets instability. Numerical solution of Eq. (\ref{NLSE2}) with
these initially truncated LLBs shows exponentially growing harmonic oscillations on their top. For the three-dimensional case and with $M=10$,
e.g., Fig. \ref{instability1}(a) shows the axial intensity $|\tilde A(0,\xi)|^2$ for perturbed LLBs with $\gamma=0$ (lossless limit) and with
$\gamma=0.1$, and Fig. \ref{instability1}(b) shows, in logarithmic scale, the difference $|\tilde A(0,\xi)|^2-1$ with respect to the unperturbed
propagation. For each specific LLB, the exponential gain $-\kappa_I$ and oscillation frequency $\kappa_R$ is seen to be independent of the initial
weak perturbation (different truncation radii). The weaker the input perturbation (the larger the truncation radius), the cleaner the exponentially
growing oscillations. As seen in Fig. \ref{instability1}(a), more LLBs preserve its shape for longer propagation distances than less LLBs. This
is because the instability gain, i.e., the slope in the logarithmic scale in Fig. \ref{instability1}(b), diminishes with increasing losses. For each
LLB (value of $\gamma$), fitting $|\tilde A(0,\xi)|^2-1$ with the same quantity given by Eq. (\ref{PERT}) in the axial region of small perturbation,
allows to obtain the gain $\kappa_I$ of the dominant unstable mode and its axial oscillation frequency $\kappa_R$. They are represented in
Figs. \ref{instability2}(a) and (b) as functions of $\gamma$ for the three-dimensional LLBs with $M=10$. Numerically, it is very difficult to prolong
the curves $\kappa_I$--$\gamma$ and $\kappa_R$--$\gamma$ up to the limit $\gamma_{\rm max}$ of the most LLB
because LLBs become less and less localized, and the truncation radii must then be extremely large for the initial perturbation to remain weak.
Nevertheless, Fig. \ref{instability2} evidences the stabilizing role of NLLs. The gain decreases monotonically from its highest value
in the limit of negligible losses ($\gamma=0$) down to negligible or zero when the limit of the most LLB is approached. This result confirms that
the most LLB is the most stable light bullet without a conical structure, and explains that it acts as the attractor in the self-focusing and collapse
arrested by NLLs of input pulses without a conical structure.

\begin{figure}
\begin{center}
\includegraphics[width=4cm]{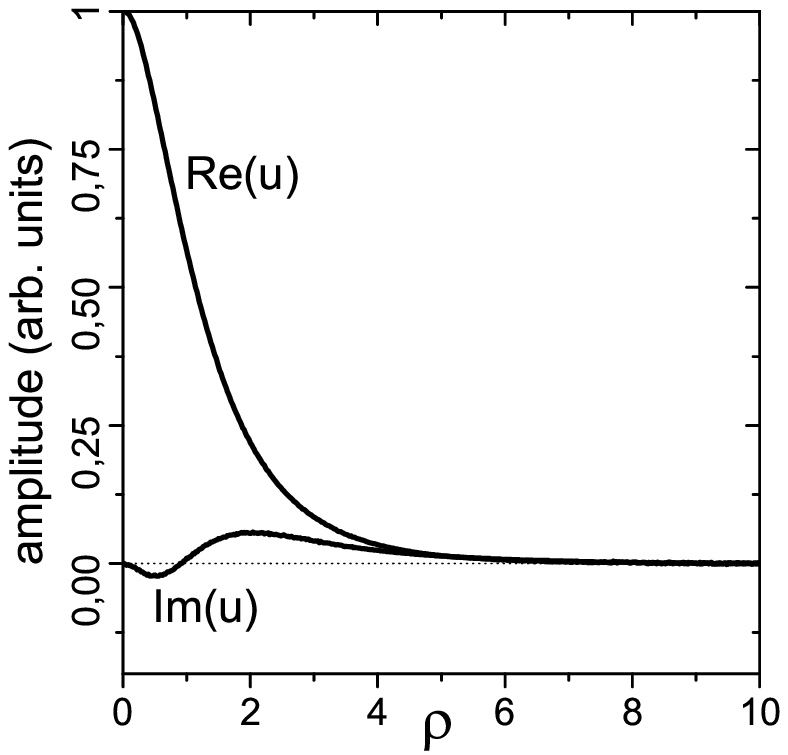}\includegraphics[width=4cm]{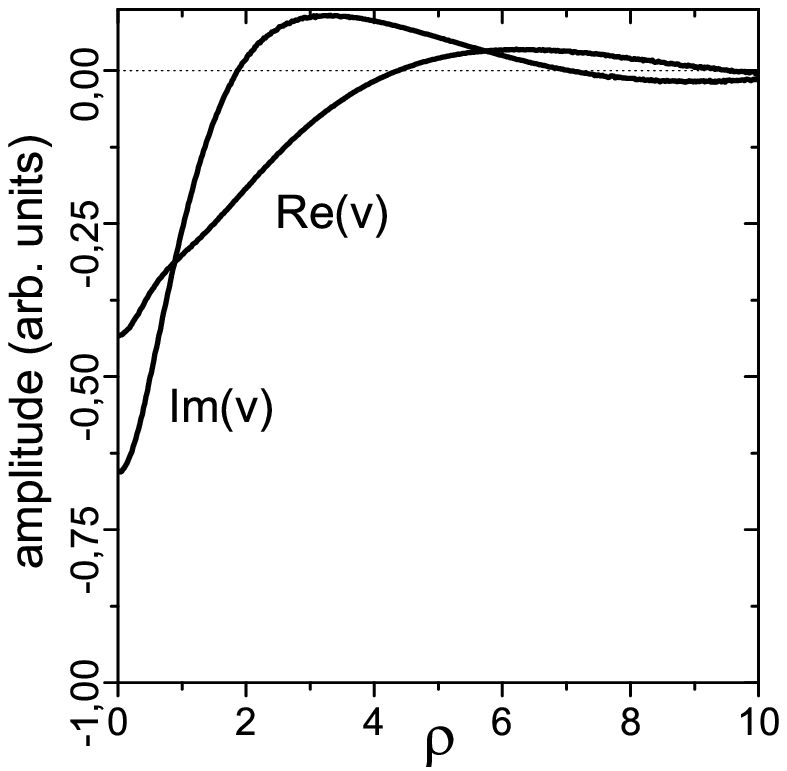}
\caption{\label{instability3} Parts (a) $u$ and (b) $v$ of the dominant unstable eigenmode of the three-dimensional LLB in media
with $M=10$ and $\gamma=0.1$. Normalization is such that ${\rm Im} (u)=0$.}
\end{center}
\end{figure}

Once the eigenvalue of the dominant unstable mode of each LLB is found, it is not difficult to evaluate its radial shape. Setting
Eq. (\ref{PERT}) with the propagated fields $\tilde A(\rho,\xi_1)$ and $\tilde A(\rho,\xi_2)$ at two distances in the region of small perturbation,
obtained from the NLSE (\ref{NLSE2}), the radial shape of the unstable mode can be evaluated from
\begin{equation}\label{UV}
[u(\rho),v(\rho)]=\frac{g_1e^{\kappa_I\xi_1}-g_2e^{\kappa_I\xi_2}e^{\mp\kappa_R(\xi_1-\xi_2)}}{e^{\pm\kappa_R\xi_1}
[1-e^{2i\kappa_R(\xi_1-\xi_2)}]}\, ,
\end{equation}
where the upper (lower) plus-minus signs stand for $u$ ($v$), and $g_i=\tilde A(\rho,\xi_i)-\tilde a(\rho)$, $i=1,2$. Except for small
fluctuations, the unstable mode obtained from Eq. (\ref{UV}) is independent of the input weak perturbation and of the couple of distances
chosen, which supports the consistency of our instability analysis.  As an example, Fig. \ref{instability3} shows the most unstable mode growing
on the top of the three-dimensional LLB with $M=10$ and $\gamma=0.1$.

\section{Conclusions \label{CONCLUSIONS}}

Summarizing, we have reviewed the properties of purely nonlinear localized waves sustained by a dynamic
equilibrium between self-focusing and nonlinear losses. Their finite-energy versions preserve light bullet behavior well-beyond the diffraction or
dispersion distances, and they rebuild after obstacles. Lossy light bullets are essentially multidimensional waves because the replenishment
mechanism from the energy reservoir is based solely on the trend toward collapse, which does not exist with one dimension. There is a preferential
lossy light bullet with maximum intensity and losses, and defined solely by the optical properties of the medium. This is the most
stable, non-conical localized wave sustained by a medium with self-focusing nonlinearity and nonlinear losses, and as such acts an as attractor in
the self-focusing dynamics with nonlinear losses of non-conical wave packets, as an input Gaussian wave packet.


\begin{thebibliography}{99}
\bibitem{NOTA} This peak power is slightly higher than the power $P_{\rm
cr}=2\lambda_0^2/[4\pi n(\omega_0)n_2]$ for spatial self-focusing.
\end{thebibliography}
\end{document}